\input harvmac

\newcount\figno
\figno=0
\def\fig#1#2#3{
\par\begingroup\parindent=0pt\leftskip=1cm\rightskip=1cm
\parindent=0pt
\baselineskip=11pt
\global\advance\figno by 1
\midinsert
\epsfxsize=#3
\centerline{\epsfbox{#2}}
\vskip 12pt
{\bf Fig. \the\figno:} #1\par
\endinsert\endgroup\par
}
\def\figlabel#1{\xdef#1{\the\figno}}
\def\encadremath#1{\vbox{\hrule\hbox{\vrule\kern8pt\vbox{\kern8pt
\hbox{$\displaystyle #1$}\kern8pt}
\kern8pt\vrule}\hrule}}

\Title{\vbox{\baselineskip12pt
\hbox{hep-th/0009019}
\hbox{TIFR-TH/00-46}
\hbox{OHSTPY-HEP-T-00-016}
\hbox{BROWN-HET-1231}
}}
{\vbox{\centerline{Vibration modes of giant gravitons}}}
\smallskip
\centerline{Sumit R. Das},
\smallskip
\centerline{{\it Tata Institute of Fundamental Research}}
\centerline{\it Homi Bhabha Road, Mumbai 400 005, INDIA}
\bigskip
\centerline{Antal Jevicki}
\smallskip
\centerline{{\it Department of Physics, Brown University}}
\centerline{\it Providence, RI 02192, U.S.A.}
\bigskip
\centerline{Samir D. Mathur}
\smallskip
\centerline{{\it Department of Physics, The Ohio State University}}
\centerline{\it Columbus, OH. 43210,  U.S.A.}
\bigskip

\medskip
\bigskip
\bigskip

\noindent

\def\p{\partial}

We examine the spectrum of small vibrations of giant gravitons, when
the gravitons expand in the anti-de-Sitter space and when they expand
on the sphere.  For any given angular harmonic, the modes are found to
have frequencies related to the curvature length scale of the
background; these frequencies are independent of radius
(and hence angular momentum) of the brane itself. This implies that
the holographic dual theory must have, in a given R charge sector,
low-lying non-BPS excitations with level spacings independent of
the R charge.
\bigskip
\bigskip

\Date{September, 2000}

\newsec{Introduction}

One of the interesting recent discoveries in string theory is the
fact that objects that are naively pointlike may in fact be
extended branes. The
Myers effect \ref\myers{R. Myers, JHEP 9912 (1999) 022,
hep-th/9910053.} implies that in the presence of field strengths of the gauge
fields in string theory, certain branes can expand into  higher
dimensional branes.   McGreevy, Susskind and Toumbas
\ref\giant{J. McGreevy, L. Susskind and N. Toumbas, JHEP 0006 (2000)
008, hep-th/0003075.}  considered the behavior of gravitons in the
near horizon geometries produced by branes. They argued that while
these gravitons may appear pointlike at low angular momentum, one
could (at a classical level) find extended brane states - called
``giant gravitons'' - carrying the same energy and angular momentum,
and these extended objects could be the correct picture of the
graviton at large angular momenta.  For $AdS_m \times S^n$ spacetimes
these can be a $(n-2)$-brane wrapped around a $S^{n-2}$ contained in
$S^n$ \giant\ or a $(m-2)$-brane wrapped around a $S^{m-2}$ contained
in $AdS_m$ \ref\myersrecent{M. Grisaru, R. Myers and O. Tafjord,
hep-th/0008015.}\ref\hashimoto{ A.  Hashimoto, S. Hirano and
N. Itzhaki, hep-th/0008016.}. The equilibrium configuration consists of
the brane rotating rigidly without a change of its size, and saturates
a BPS bound for the energy $E$ for a given angular momentum $P_\phi$,
$E \geq {P_\phi/L}$ where $L$ is the radius of the $S^n$
\ref\djm{S.R. Das, A. Jevicki and S.D.  Mathur, hep-th/0008088.}.
The angular velocity is {\it independent} of the angular momentum and
depends only on $L$.  When embedded in
a supersymmetric theory these configurations respect half of the
supersymmetries of the background \myersrecent,\hashimoto\ and
therefore the BPS bound follows from supersymmetry.  The structure of
the final quantum state representing the gravitons is not yet clear,
as there can be tunneling between these various classical
configurations \myersrecent,\hashimoto,\djm, and also to multiple
brane states with the same quantum numbers \djm. Interestingly, brane
states with energies equal to gravitons with the same angular momentum
have been also found to occur in spacetimes other than $AdS_m \times
S^n$
\ref\dtv{S.R. Das, S.P. Trivedi and S. Vaidya, hep-th/0008203}. In some
of these cases there are also extended branes with nonzero $D0$ brane
charge which are degenerate with pointlike $D0$ branes and in this
situation the phenomenon may be understood
quantitatively in terms of the magnetic analog
of Myers' effect \dtv.

An important issue related to these `giant graviton' states is whether
they can provide an interpretation of the stringy exclusion principle
\ref\exclusion{J. Maldacena and A. Strominger, JHEP 9812 (1998) 005,
hep-th/9804085} in terms of the limits on the allowed size of the
expanded brane configurations \giant.  It has been argued that the
stringy exclusion principle implies that the supergravity should live
on a noncommutative spacetime, e.g. quantum deformations of $AdS_m
\times S^n$ \ref\ramjev{A. Jevicki and S.  Ramgoolam, JHEP 9904 (1999)
032, hep-th/9902059; P. Ho, S. Ramgoolam and R.  Tatar, Nucl. Phys.  B
573 (2000) 364, hep-th/9907145; A. Jevicki, M. Mihailescu and
S. Ramgoolam, hep-th/0006239; A. Jevicki, M. Mihailescu and
S. Ramgoolam, hep-th/0008186.}.  Structures similar to expanded branes
can also be obtained using the non-abelian interactions of
multi-particle systems in string theory \ref\kt{ D.  Kabat and
W. Taylor, Adv. Theor. Math. Phys. 2 (1998) 181, hep-th/9711078.}
\ref\bv{M.Berkooz and H.Verlinde, JHEP 9911 (1999)037,
hep-th/9907100.}\ref\jy{A.Jevicki and T.Yoneya,Nucl.Phys.B 535 (1998)
335, hep-th/ 9805069.}\ref\bjl{D.Berenstein, V.Jejjala and R.Leigh,
hep-th/0005087.}\ref\bhp{C.Bachas,J.Hoppe and
B.Piolline,hep-th/0007067.} where a nonocommutative structure arises.
Indeed for the giant gravitons there are hints of such a
noncommutativity of space emerging \djm,\ref\holi{P. Ho and M. Li,
hep-th/0004072.}.

Extended objects possess a set of low energy excitations arising from
small vibrations about their equilibrium configuration.  Such modes
have been very important in string theory in the study of black
holes. If we regard a black hole as a point singularity surrounded by
empty space then we cannot account for its microscopic degrees of
freedom.  But a string theory description of charged holes replaces
the pointlike matter by extended branes, and vibration modes of these
branes can be used to derive the microscopic entropy
\ref\entropy{L. Susskind: 1993,
hep-th 9309145;  A. Sen: 1995,
{\it Nucl. Phys. } {\bf B440},
p 421; 1995: {\it Mod. Phys. Lett.} {\bf A10} p2081;  A. Strominger
and C. Vafa: 1996,
{\it Phys. Lett.} {\bf B379},
p 99; C. Callan and J. Maldacena: 1996,
{\it Nucl. Phys.} {\bf B472},
p 591.} and unitary low energy Hawking radiation for
these black holes \ref\radiation{S.R. Das and S.D. Mathur: 1996,
{\it Nucl. Phys.} {\bf B478},
p 561;  S.R. Das and S.D. Mathur: 1996,
{\it Nucl. Phys.} {\bf B482}, p 153;  A. Dhar. G. Mandal and S.
Wadia: 1996, {\it Phys. Lett.}  {\bf B388}, 51; J.
Maldacena and A. Strominger: 1997,
      {\it Phys. Rev. }  {\bf D55}, 861.}.

There are several reasons why we would like to study the vibration
spectrum of giant gravitons.  First, if there is a family of
solutions with the same energy and angular momentum, then it would
show up as a corresponding mode of the vibration spectrum. Thus
looking at such modes would provide a way of checking whether the
spherical brane ansatz used in \giant\ captures all the states that
are classically BPS. Secondly, the branes in AdS space with angular
momentum greater than the exclusion principle bound pose a puzzle
\myersrecent ;   we would like to check if these
configurations are unstable to some harmonic of vibration (we do not
in fact find such an instability). Thirdly, such vibration modes give
a closely spaced set of low energy motions of the graviton that could
be excited in interactions - thus these modes appear to be essential
to finding the inclusive cross section for the interaction of two
gravitons.  In particular, one finds the following aspect of the
stringy exclusion principle
in the $D1-D5$ system. One can compute the 3-point function of chiral
primaries in the boundary CFT,
for operators that are dual to supergravity modes in the bulk theory
\ref\three{A. Jevicki, M. Mihailescu and
S. Ramgoolam, Nucl. Phys. B577 (2000) 47, hep-th/9907144
and hep-th/0006239}
\ref\luninmathur{O. Lunin and S. D. Mathur, hep-th/0006196}.  It was
found in \luninmathur\ that the three point function starts dropping
significantly below the naive supergravity expectation when the
angular momenta are in fact  much smaller
than the allowed upper bound.  If the giant graviton picture is
right, then it could shed some light on this
drop in the 3-point function.

In this paper we study the vibration spectrum of giant gravitons in
$AdS_m \times S^n$. We focus on the excitations arising from motion of
the branes in spacetime, and thus do not consider here the excitations
that arise from fermionic modes, or from any gauge fields that may
live on the brane representing the giant graviton.  The embedding
geometry has the structure of an anti-de-Sitter (AdS) spacetime times
a sphere. We consider both the case where the brane expands in the AdS
spacetime, and the case where it expands on the sphere. In the former
case the radius of the brane can be much larger than the curvature
scale of the embedding spacetime, while in the latter case the radius
of the brane is limited by the size of the sphere.

One of the questions that we shall focus on is the scale which gives
the vibration frequencies.  There are several scales in the problem:
the microscopic string scale or planck scale, the radius of the brane,
and the curvature scale of the spacetime (the AdS spacetime and the
sphere have curvatures of the same order).  A priori the frequency of
vibration can be any combination of these scales which has the right
dimension.  Since we are extremising the classical action we will not
encounter the string or planck scales, but it is not immediately
obvious which of the other scales should give the vibration
frequencies.

We find that the frequencies are all real and, somewhat surprisingly,
{\it indpendent of the size of brane}. They depend only on the $AdS$
scale ${\tilde L}$ and the radius of the sphere $L$. The system thus
has  low energy excitations with spacings
independent of the angular momentum of the
brane. This gives a prediction for the holographic
dual:  for a given R-charge sector of the dual
theory, one should find a non-BPS spectrum of
excitations with level spacings independent of the R
charge.

The plan of this paper is the following. In section 2 we set up
notation and describe the equilibrium configurations about which we
will study the vibrations. In section 3 we find the spectrum for the
branes that expand in the AdS spacetime, and in section 4 we repeat
this calculation for branes that expand on the sphere.  Section 5
studies some aspects of the frequencies and
section 6 is a general discussion of the implications  of the spectrum.

\newsec{The equilibrium configurations}

We will for the most part follow the notation of \myersrecent .  The
spacetime will have the form $AdS_m\times S^n$.  In particular we have
the $D=11$ supergravity backgrounds with $(m,n)=(4,7)$ and
$(m,n)=(7,4)$, and the $D=10$ background of IIB supergravity with
$(m,n)=(5,5)$.  In the first two cases the graviton can expand into
2-branes and 5-branes of M-theory, while in the last case it can
expand into the D-3 brane of type IIB string theory. In addition we
have $(m,n)=(3,3)$ for the cases $AdS_3\times S^3\times M^4$, though
this case is more involved since since the 1-branes can be various
combinations of strings and 5-branes wrapped on the compact 4-manifold
($T^4$ or K3).

\subsec{Branes expanding in AdS spacetime}

We will find it convenient to use slightly different coordinates when
looking at  branes expanding in the  AdS and when looking at branes
expanding
on the sphere. The AdS space and the sphere are orthogonal to each other
\eqn\yone{ds^2=ds^2_{AdS}+ds^2_{sph} }
(In the case of $AdS_3\times S^3\times M^4$ the $M^4$ is orthogonal
to the other two components as well.)

For branes in AdS, we write the AdS metric in global coordinates as
\eqn\one{ds^2_{AdS}=-(1+{r^2\over \tilde L^2})dt^2+{dr^2\over
(1+{r^2\over \tilde L^2})} +r^2d\Omega_{m-2}^2}
We can further write the metric $d\Omega_{m-2}^2$ as
\eqn\two{d\Omega_{m-2}^2=d\alpha_1^2+\sin^2\alpha_1(d\alpha_2^2+\sin^2
\alpha_2(\dots +\sin^2\alpha_{m-3}d\alpha_{m-2}^2))}

The sphere $S^n$ has radius $L$.  We write
\eqn\three{ds^2_{sph}=L^2d\Omega_{n}^2}
where $d\Omega_{n}^2$ is the metric on the unit $n$ sphere. We
further describe this unit sphere by using coordinates $z_1, z_2,
y_i, i=1\dots n-1$
\eqn\four{z_1^2+z_2^2+ y_1^2+\dots y_{n-1}^2=1}
\eqn\five{z_1^2+z_2^2=\cos^2\theta, ~~~~~~\sum_{k=1}^{n-1} y_k^2=\sin^2\theta}
\eqn\six{z_1=\cos\theta\cos\phi, ~~~~~~z_2=\cos\theta\sin\phi}
A complete set of cordinates for $S^n$ is $ \phi, y_k$. In these
coordinates the metric for $S^n$ is
\eqn\sixp{ds_{sph}^2=L^2[(1-\sum_k y_k^2)
d\phi^2+(\delta_{ij}+{y_iy_j\over 1-\sum_ky_k^2}) dy_idy_j] }

The brane in its unexcited configuration moves at $\theta=0$, which
gives $y_k=0$. Further,
\eqn\seven{\phi=\omega_0 t, ~~~~\omega_0={1\over L}}
     \eqn\eight{r=r_0}
The worldsheet extends for all $t$, and wraps the $m-2$ sphere in
\one , covering all values of the coordinates $\alpha_i, i=1\dots
m-2$. The angular momentum $P_\phi$ is given by
\eqn\sevenj{P_\phi = {\tilde N} ({r_0 \over {\tilde L}})^{m-3}}
where
\eqn\sevenk{{\tilde N} = V_{m-2} T_{m-2} L {\tilde L}^{m-2}}
and $V_{m-2}$ is the volume of the unit $S^{m-2}$ and $T_{m-2}$ is the
tension of the $(m-2)$-brane. The energy of this state is given by
\eqn\sevenl{ E = {P_\phi \over L}}

\subsec{Branes expanding on the sphere}

For this case we must essentially interchange the roles of the AdS
and sphere metrics in the above discussion, and so we adopt the
following
coordinates.  We describe an $AdS_m$ spacetime of unit radius through
\eqn\nine{-(u_1^2+u_2^2)+(v_1^2+\dots v_{n-1}^2)=-1}
\eqn\ten{u_1^2+u_2^2=\cosh^2\mu, ~~~~~~\sum_{k=1}^{m-1} v_k^2=\sinh^2\mu}
\eqn\el{u_1=\cosh\mu\cos{t\over \tilde L},
~~~~~u_2=\cosh\mu\sin{t\over\tilde L} }
A complete set of coordinates for the AdS space is $t, v_k$. In
these coordinates the metric for $AdS_m$ is
\eqn\tw{ds^2_{AdS}=-(1+\sum_k v_k^2)dt^2+\tilde
L^2(\delta_{ij}+{v_iv_j\over 1+\sum_k v_k^2}) dv_idv_j }
(We pass to the covering space where $t$ runs over $(-\infty,
\infty)$ instead of $(0, 2\pi)$ to recover the complete AdS
spacetime.)

The metric of the sphere $S^n$ is
\eqn\thir{ds_{sph}^2=L^2[d\theta^2+\cos^2\theta d\phi^2+\sin^2\theta
d\Omega_{n-2}^2]}
Here $\theta, \phi$ are the same coordinates as introduced in \five ,
\six\ and where we can write
\eqn\fourt{d\Omega_{n-2}^2=d\chi_1^2+\sin^2\chi_1(d\chi_2^2+\sin^2\chi
_2(\dots +\sin^2\chi_{n-3}d\chi_{n-2}^2))}
This time the unexcited brane will have a fixed value of $\theta$
between $0$ and $\pi$. We write
\eqn\fift{\sin\theta={q\over L}}
Then the metric on $S^n$ becomes
\eqn\sixt{ds_{sph}^2=L^2 (1-{q^2\over L^2}) d\phi^2 + {dq^2\over
1-{q^2\over L^2}} + q^2d\Omega_{n-2}^2}
which resembles the form \one\ of the AdS metric.  The brane has a
radius determined by
\eqn\sevent{q=q_0.}
      $\phi (t)$ is again given by \seven , and the brane extends
over the coordinate
$t$, and over the coordinates $\chi_i$ in \fourt . The angular momentum
is now
\eqn\seventj{P_\phi = N({q_0 \over L})^{n-3}}
where
\eqn\seventk{ N = L^{n-1} V_{n-2}T_{n-2}}
is the quantized flux of the $n$-form field strength. The energy is
still given by \sevenl.

The length scales of the AdS  and the sphere are related by
\eqn\nint{{\tilde L\over L}={m-1\over n-1}}

\newsec{Vibration spectrum - branes in AdS}

The action for the brane is
\eqn\tone{S=S_{DBI}+S_{CS}}
where $S_{DBI}$ is the Dirac-Born-Infeld action, and $S_{CS}$ is the
Chern-Simons term.

For a $p$ brane, $S_{CS}$ is
\eqn\ttwo{S_{CS}=T_p\int P[A^{(p+1)}]}
where $P$ denotes the pullback of the $p+1$ form gauge potential onto
the brane worldvolume.  We will consider a $m-2$ brane. The gauge
potential in
spacetime gives a constant field strength on $AdS_m$ and its dual
(constant) field strength on $S^n$.  The potential on $AdS_m$
is\foot{We choose
signs of the gauge potentials that are different from those in \myersrecent.}
\eqn\tfour{A^{(m-1)}_{t\alpha_1\dots\alpha_{m-2}}={r^{m-1}\over
\tilde L} \sqrt{g_\alpha} }
where $\sqrt{g_\alpha}d\alpha_1\dots d\alpha_{m-2}$ is the volume
element on the $m-2$ sphere. Thus the field strength is
\eqn\tfive{F^{(m)}_{rt\alpha_1\dots\alpha_{m-2}}=(m-1){r^{m-2}\over
\tilde L} \sqrt{g_\alpha} }
In a local orthonormal frame the value of $F$ is
\eqn\tsix{F^{(m)}_{\hat r\hat
t\hat\alpha_1\dots\hat\alpha_{m-2}}={(m-1)\over \tilde L}={\rm
constant}  }

The potential on the sphere is
\eqn\qone{A^{(n-1)}_{\phi\chi_1\dots\chi_{n-2}}=L^{n-1}\sin^{n-1}\theta
\sqrt{g_\chi}={q^{n-1}}\sqrt{g_\chi}}
where $\sqrt{g_\chi}d\chi_1\dots d\chi_{n-2}$ is the volume element
on the $n-2$ sphere \fourt , and $q$ was defined in  \fift . The
field strength is
\eqn\qel{F_{q\phi\chi_1\dots\chi_{n-2}}^{(n)}=(n-1)q^{n-2}\sqrt{g_\chi}}
In a unit orthonormal frame we get
\eqn\qtw{F_{\hat q\hat \phi\hat \chi_1\dots\hat
\chi_{n-2}}^{(n)}={(n-1)\over L}}

Using \nint\ we see that \tsix\ and \qtw\ are dual forms.

\subsec{The action}

The  configuration of the brane is described by giving the spacetime
coordinates as a function of the worldsheet coordinates $\tau,
\sigma_1,\dots
\sigma_{m-2}$. We choose the static gauge, where
\eqn\qfour{t=\tau}
\eqn\qfive{\alpha_i=\sigma_i, ~~~~i=1\dots m-2}
The remaining coordinates are given by
\eqn\qsix{r=r_0+\epsilon ~\delta r(\tau, \sigma_1,\dots \sigma_{m-2})}
\eqn\qseven{\phi=\omega_0 \tau +\epsilon~\delta\phi(\tau,
\sigma_1,\dots \sigma_{m-2})}
\eqn\qeight{y_k=\epsilon ~\delta y_k(\tau, \sigma_1,\dots
\sigma_{m-2}), ~~~~k=1\dots n-1}
Recall that we are describing $AdS_m$ by $t, r, \alpha_1,\dots
\alpha_{m-2}$ and $S^n$ by $\phi$,  $y_1,\dots y_{n-1}$.

Let $G_{MN}$ be the metric of $AdS_m\times S^n$, and let $g_{ij}$ be
the induced metric on the brane
\eqn\wtwo{g_{ij}={\p X^M\over \p \xi^i}{\p X^N\over \p\xi^j}G_{MN} }
where $X^M$ are coordinates on $AdS_m\times S^n$ and $\xi^i$ are
coordinates on the worldsheet.
The DBI action is
\eqn\wonw{S_{DBI}=-T_{m-2}\int \sqrt{-g}d\tau d\sigma_1\dots d\sigma_{m-2}}

The Chern-Simmons term has two possible contributions, from the two
nonvanishing components \tfour\ and \qone ; we call them $S_{CS1}$ and
$S_{CS2}$ repectively. We are interested only in the action to
quadratic order in the fluctuations, and  we will find that $S_{CS2}$
does not contribute if
$n>3$,
since its contribution becomes higher order than quadratic in the
fluctuation. Thus the case $n=3$ will have to be treated separately
when computing
the vibration frequencies. (We do not consider $n<3$ or $m<3$.)

The pullback of the gauge field is
\eqn\qthree{P[A^{(m-1)}]_{\tau\sigma_1\dots\sigma_{m-2}}=A^{(m-1)}_{M_
1\dots M_{m-1}}{\p X^{M_1}\over \p\tau }{\p X^{M_2}\over \p
\sigma_1}\dots {\p X^{M_{m-1}}\over \p \sigma_{m-2} }}

     Using  \qfour\ - \qeight\ we get
\eqn\qtwo{\eqalign{S_{CS1}&=T_{m-2}\int
A^{(m-1)}_{t\alpha_1\dots\alpha_{m-2}} d\tau d\sigma_1\dots
d\sigma_{m-2}\cr
&=T_{m-2} \int {1\over \tilde L}(r_0+\epsilon \delta r(\tau,
\sigma_1,\dots \sigma_{m-2}))^{m-1}\sqrt{g_\sigma}d\tau d\sigma_1\dots
d\sigma_{m-2}\cr}}
where $\sqrt{g_{\sigma}}d\sigma_1\dots d\sigma_{m-2}$ is the volume
element on a constant $\tau$ hypersurface on the worldsheet.

For $S_{CS2}$ we write down only the terms that contribute to lowest
nonzero order in $\epsilon$.  For small values of  $q, y_k$ we see
from \qtw\
that we can write the gauge potential on $S^n$ as
\eqn\ten{A^{(n-1)}_{\phi y_2\dots y_{n-1}}\approx  L^{n-1}{(n-1)}y_1 }
Then we get
\eqn\qthir{\eqalign{S_{CS2}&\approx T_{m-2}L^{n-1}\int
(n-1)y_1{\p\phi\over \p\tau}{1\over (m-2)!}\epsilon_{i_1\dots
i_{m-2}}{\p y_2\over \p
\sigma_{i_1}}\dots {\p y_{m-1}\over\p\sigma_{i_{m-2}}} d\tau
d\sigma_1\dots d\sigma_{m-2}\cr
&\approx T_{m-2}L^{n-1}~\omega_0 (n-1)\int y_1{1\over
(m-2)!}\epsilon_{i_1\dots i_{m-2}} {\p y_2\over \p
\sigma_{i_1}}\dots {\p y_{m-1}\over\p\sigma_{i_{m-2}}}d\tau
d\sigma_1\dots d\sigma_{m-2}\cr}}
For this to be nonvanishing, we need $n= m$. But further, the order
of this term is $m-1$ in the perturbation, so for $m>3$ it is not
relevant for the
linearised perturbation analysis. It {\it would} be relevant for
$AdS_3\times S^3$, though this case is more involved since the
1-branes can be various
combinations of strings and 5-branes wrapped on the compact
4-manifold ($T^4$ ot K3). We will analyze these aspects of the
$AdS_3\times S^3$ case elsewhere, but for completeness work out here
the frequencies that
follow  from an action of the form $S=-T_1\int \sqrt{-g}+T_1\int
P[A^{(2)}]$.

\subsec{Linearised equations  for $m>3$}

First we look at the linear term in $\epsilon $ in the action \tone .
A straightforward calculation gives
\eqn\wfive{\eqalign{S_{DBI}\pm S_{CS1}=&-\epsilon~T_{m-2}~\int d\tau
d\sigma_1\dots d\sigma_{m-2}\sqrt{g_\sigma}\cr
&{r_0^{m-3}\over \tilde
L}[\{{(m-1)r_0^2+(m-2)\tilde L^2(1-L^2\omega_0^2)\over
\sqrt{r_0^2+\tilde L^2(1-L^2\omega_0^2)}}\mp r_0(m-1)\}\delta r \cr
&-{L^2\tilde L^2\omega_0r_0\over
\sqrt{r_0^2+\tilde L^2(1-L^2\omega_0^2)}}{\p\delta
\phi\over
\p\tau}]\cr }}
As is clear from this formula, the action of the equilibrium configuration
vanishes.

The coefficient of the term ${\p\delta
\phi\over
\p\tau}$ is a constant, and so this term gives no contribution to the
variation of the action with fixed boundary values. The coefficient
of the term
$\delta r$ vanishes if we take
\eqn\wseven{\omega_0=\pm {1\over L} }
and the $+$ sign on the LHS of \wfive .  We choose the positive sign
in \wseven\ for concreteness; the frequncies we find are independent
of this
choice.

Looking at the linear order variation \wfive\ we see that we also get
the coefficient of $\delta r$ to vanish if we choose
\eqn\tone{\omega_0^2={1\over L^2}[1+{r_0^2\over \tilde
L^2}{(m-1)(m-3)\over (m-2)^2}]}
These solutions should correspond to  the maxima of the potential  in
\myersrecent  \hashimoto , and thus describe unstable configurations.
We will
not consider perturbations around these configurations.

With the choices given in \wseven\ (and immediately following that
equation) we find that the zeroth order term in
$\epsilon$ vanishes, while the second order term in
$\epsilon$ is
\eqn\weight{\eqalign{S=&\epsilon^2~T_{m-2}~r_0^{m-3}\int d\tau
d\sigma_1\dots d\sigma_{m-2}\sqrt{g_\sigma}\cr
&[{\tilde L^3\over 2 (r_0^2+\tilde L^2)}   ({\p\delta r\over
\p\tau})^2 -{\tilde L\over 2(r_0^2+\tilde L^2)} {\p\delta r\over
\p\sigma_i}{\p\delta r\over
\p\sigma_i}g^{\sigma_i\sigma_j} \cr
     &+{L^2\tilde L(r_0^2+\tilde L^2)\over 2 r_0^2}({\p\delta\phi\over
\p\tau})^2 -{L^2(r_0^2+\tilde L^2)\over 2\tilde Lr_0^2}
{\p\delta \phi\over \p\sigma_i}{\p\delta \phi\over
\p\sigma_i}g^{\sigma_i\sigma_j}  \cr
&+{L\tilde L(m-3)\over r_0}\delta r  {\p\delta\phi\over \p\tau}  \cr
&+{L^2\tilde L\over 2} {\p\delta y_k\over \p\tau}{\p\delta y_k\over
\p\tau} -{L^2\over 2\tilde L}
{\p\delta y_k\over \p\sigma_i}{\p\delta y_k\over
\p\sigma_i}g^{\sigma_i\sigma_j} -{\tilde L\over 2}y_ky_k] \cr}}

Let $Y_l$ be spherical harmonics on the unit $m-2$ sphere
\eqn\eone{g^{\sigma_i\sigma_j}{\p\over \p\sigma_i}{\p\over
\p\sigma_j}Y_l(\sigma_1\dots \sigma_{m-2})=-Q_lY_l(\sigma_1\dots
\sigma_{m-2})}
For example, on a 2-sphere we have $Q_l=l(l+1)$.

We expand the perturbations as
\eqn\etwo{\eqalign{\delta r(\tau, \sigma_1\dots\sigma_{m-2})=&
\tilde\delta r e^{-i\omega\tau}Y_l(\sigma_1\dots \sigma_{m-2})\cr
\delta \phi(\tau, \sigma_1\dots\sigma_{m-2})=& \tilde\delta \phi
e^{-i\omega\tau}Y_l(\sigma_1\dots \sigma_{m-2})\cr
\delta y_k(\tau, \sigma_1\dots\sigma_{m-2})=&\tilde \delta y_k
e^{-i\omega\tau}Y_l(\sigma_1\dots \sigma_{m-2})\cr}}

We see that the $\delta y_k$ perturbations decouple from $\delta r,
\delta \phi$, and have frequencies given by
\eqn\efour{\omega^2={Q_l\over \tilde L^2}+{1\over  L^2}}

The $\delta r, \delta\phi$ perturbations are seen to be coupled. The
resulting frequencies are given by the equation
\eqn\efive{\pmatrix{{\tilde L\over (r_0^2+\tilde
L^2)}(-\omega^2\tilde L^2+Q_l)& i\omega (m-3) {L\tilde L\over r_0}\cr
-i\omega(m-3){L\tilde L\over r_0}&{L^2(r_0^2+\tilde L^2)\over \tilde
L r_0^2}(-\omega^2\tilde L^2+Q_l)\cr}\pmatrix{\tilde\delta r\cr
\tilde\delta\phi\cr}~=~0}
which yields
\eqn\esix{\omega^2_{\pm}={1\over \tilde L^2}[Q_l+{(m-3)^2\over 2}\pm
(m-3)\sqrt{Q_l+{(m-3)^2\over 4}}]}

\subsec{The case $n=3, m=3$}

In this case   the $m-2$ sphere is just a circle, which is
parametrized by only one coordinate $\sigma_1\equiv \sigma$.  Thus we
have
\eqn\ethir{Y_l(\sigma)=e^{il\sigma}, ~~~~Q_l=l^2}
    From \qthir\  we
get an  additional contribution to the action
\eqn\eten{S_{CS2}=2T_1L^2\omega_0\int y_1 {\p y_2\over \p\sigma} d\tau d\sigma}

The $\delta r, \delta\phi$ perturbations are unaffected by $S_{CS2}$,
and so we get the equation \efive\ with $m=3$. Thus the $\delta r$
and $\delta
\phi$ perturbations decouple, and each has a frequency given by
\eqn\etw{\omega^2={Q_l\over \tilde L^2}={l^2\over \tilde L^2}}

There are two $y$ coordinates, $y_1$ and $y_2$. The frequencies of
their fluctuations are given by
\eqn\eel{\pmatrix{{L^2\over \tilde L}(-\omega^2\tilde L^2+l^2)+\tilde
L& -2iLl\cr
2iLl & {L^2\over \tilde L}(-\omega^2\tilde L^2+l^2)+\tilde
L\cr}\pmatrix{\tilde\delta y_1\cr \tilde\delta y_2\cr}=0}
which gives the frequencies
\eqn\eeight{\omega^2_\pm={1\over \tilde L^2}(l\pm {\tilde L\over
L})^2={1\over \tilde L^2}(l\pm 1)^2}
where in the last step we have used the fact that by \nint\ $L=\tilde
L$ in this case.

\newsec{Vibration spectrum - branes on the sphere}

In this case the brane worldsheet has dimension $n-1$, and we
describe it by coordinates $\tau, \sigma_1,\dots \sigma_{n-2}$. We
choose the static
gauge
\eqn\rone{t=\tau}
\eqn\rtwo{\chi_i=\sigma_i, ~~~~i=1\dots n-2}

The remaining coordinates are given by
\eqn\rthree{q=q_0+\epsilon~\delta q(\tau, \sigma_1,\dots\sigma_{n-2}) }
\eqn\rfour{\phi=\omega_0\tau+\epsilon~ \delta\phi(\tau,
\sigma_1\dots\sigma_{n-2})}
\eqn\rfive{v_k=\epsilon~\delta v_k(\tau, \sigma_1\dots\sigma_{n-2}),
~~~~~k=1\dots m-1}
Recall that now $AdS_m$ is described by the oordinates $t, v_1,
\dots v_{m-1}$ and $S^n$ is described by $\phi, q, \chi_1, \dots
\chi_{n-2}$.

The DBI action is
\eqn\rsix{S_{DBI}=-T_{n-2}\int \sqrt{-g} d\tau d\sigma_1\dots d\sigma_{n-2} }
The gauge field again can give two kinds of terms. This time we will
call $S_{CS1}$ the term arising from the gauge field on the sphere
$S^n$, and
$S_{CS2}$ the term from the gauge field on $AdS_m$.  Then
\eqn\rseven{\eqalign{S_{CS1}=&T_{n-2}\int
A_{\phi\chi_1\dots\chi_{n-2}}{\p \phi\over \p\tau}{\p\chi_1\over \p
\sigma_1}\dots {\p \chi_{n-2}\over
\p\sigma_{n-2}}d\tau d\sigma_1\dots d\sigma_{n-2}\cr
=& T_{n-2}\int (\omega_0+\epsilon {\p \delta\phi(\tau,
\sigma_1\dots\sigma_{n-2})\over
\p\tau})A_{\phi\chi_1\dots\chi_{n-2}}d\tau d\sigma_1\dots
d\sigma_{n-2}\cr =& T_{n-2}\int (\omega_0+\epsilon {\p
\delta\phi\over \p\tau})(q_0 + \delta q)^{n-1}\sqrt{g_\sigma}d\tau
d\sigma_1\dots
d\sigma_{n-2}\cr}}

     To compute $S_{CS2}$ we write the form of the gauge potential on
$AdS_m$  for small $v_k$
\eqn\rnine{A^{(m-1)}_{\tau v_2\dots v_{m-1}}\approx  \tilde L^{m-2}(m-1)v_1}
We write the lowest order term for $S_{CS2}$
\eqn\reight{S_{CS2}\approx T_{n-2}\tilde L^{m-2}(m-1)\int v_1{1\over
(n-2)!}\epsilon_{i_1\dots i_{n-2}}{\p v_2\over \p \sigma_{i_1}}\dots
{\p v_{n-1}\over
\p
\sigma_{n-2}}d\tau d\sigma_1\dots d\sigma_{n-2} }
Again, this term is nonvanishing only for $n=m$, and is of order
higher than quadratic in $\epsilon$ if $n>3$.

\subsec{Linearised equations for $n>3$}

Expanding the action to the  linear order term in $\epsilon$ we get
\eqn\rten{\eqalign{S_{DBI}\pm S_{CS1} &=-\epsilon~T_{n-2}\int d\tau
d\sigma_1\dots d\sigma_{n-2}\sqrt{g_\sigma} \cr
&q_0^{n-3}[\{{(n-1)q_0^2\omega_0^2+(n-2)(1-L^2\omega_0^2)\over
\sqrt{1-L^2\omega_0^2+q_0^2\omega_0^2}}\mp (n-1)q_0\omega_0 \}\delta
q\cr
& +\{{(q_0^2-L^2)q_0\omega_0\over
\sqrt{1-L^2\omega_0^2+q_0^2\omega_0^2}}\pm q_0^2\} {\p
\delta\phi\over \p \tau} ]\cr }}
The coefficient of the term ${\p\delta
\phi\over
\p\tau}$ is a constant as before, and so this term gives no
contribution to the variation of the action with fixed boundary
values. The coefficient of the
term
$\delta q$ vanishes if we take
\eqn\wsevenq{\omega_0=\pm {1\over L} }
and the $+$ sign on the LHS of \rten .  We again choose the positive
sign in \wsevenq\ for concreteness; the frequencies will be
independent of this
choice. With these choices  the zeroth order term in
$\epsilon$ vanishes, while the second order term in
$\epsilon$ is
\eqn\weightq{\eqalign{S=&\epsilon^2~T_{n-2}~q_0^{n-3}\int d\tau
d\sigma_1\dots d\sigma_{n-2}\sqrt{g_\sigma}\cr
&[{ L^3\over 2 (L^2-q_0^2)}   ({\p\delta q\over \p\tau})^2 -{ L\over
2(L^2-q_0^2)} {\p\delta q\over \p\sigma_i}{\p\delta q\over
\p\sigma_i}g^{\sigma_i\sigma_j} \cr
     &+{L^3(L^2-q_0^2)\over 2 q_0^2}({\p\delta\phi\over \p\tau})^2
-{L(L^2-q_0^2)\over 2 q_0^2}
{\p\delta \phi\over \p\sigma_i}{\p\delta \phi\over
\p\sigma_i}g^{\sigma_i\sigma_j}  \cr
&+{L^2(n-3)\over q_0}\delta q  {\p\delta\phi\over \p\tau}  \cr
&+{L\tilde L^2\over 2} {\p\delta v_k\over \p\tau}{\p\delta v_k\over
\p\tau} -{\tilde L^2\over 2 L}
{\p\delta v_k\over \p\sigma_i}{\p\delta v_k\over
\p\sigma_i}g^{\sigma_i\sigma_j} -{ L\over 2}v_kv_k] \cr}}

Let $Y_l$ be spherical harmonics on the unit $n-2$ sphere
\eqn\eoneq{g^{\sigma_i\sigma_j}{\p\over \p\sigma_i}{\p\over
\p\sigma_j}Y_l(\sigma_1\dots \sigma_{n-2})=-Q_lY_l(\sigma_1\dots
\sigma_{n-2})}
We expand the perturbations as
\eqn\etwoq{\eqalign{\delta q(\tau, \sigma_1\dots\sigma_{n-2})=&
\tilde\delta r e^{-i\omega\tau}Y_l(\sigma_1\dots \sigma_{n-2})\cr
\delta \phi(\tau, \sigma_1\dots\sigma_{n-2})=& \tilde\delta \phi
e^{-i\omega\tau}Y_l(\sigma_1\dots \sigma_{n-2})\cr
\delta v_k(\tau, \sigma_1\dots\sigma_{n-2})=&\tilde \delta v_k
e^{-i\omega\tau}Y_l(\sigma_1\dots \sigma_{n-2})\cr}}

The $v_k$ perturbations decouple from $\delta q, \delta \phi$, and
have frequencies given by
\eqn\efourq{\omega^2={Q_l\over  L^2}+{1\over  \tilde L^2}}

The $\delta q, \delta\phi$ perturbations are coupled. The resulting
frequencies are given by the equation
\eqn\efiveq{\pmatrix{{ L\over (L^2-q_0^2)}(-\omega^2 L^2+Q_l)&
i\omega (n-3) {L^2\over q_0}\cr
-i\omega(n-3){L^2\over q_0}&{L(L^2-q_0^2)\over q_0^2}(-\omega^2
L^2+Q_l)\cr}\pmatrix{\tilde\delta q\cr
\tilde\delta\phi\cr}~=~0}
which yields
\eqn\esixq{\omega^2_{\pm}={1\over L^2}[Q_l+{(n-3)^2\over 2}\pm
(n-3)\sqrt{Q_l+{(n-3)^2\over 4}}]}

\subsec{The case $n=3, m=3$}

The spherical harmonics are again given by \ethir .
We get an  additional contribution to the action
\eqn\etenq{S_{CS2}=2T_1\tilde L\int v_1 {\p v_2\over \p\sigma} d\tau d\sigma}

The $\delta r, \delta\phi$ perturbations are unaffected by $S_{CS2}$,
and  setting $n=3$ we see that they decouple from each other. Each of
these
perturbations  has a frequency given by
\eqn\etw{\omega^2={Q_l\over  L^2}={l^2\over  L^2}}

There are two $v$ coordinates, $v_1$ and $v_2$. The frequencies of
their fluctuations are given by
\eqn\eelq{\pmatrix{{\tilde L^2\over L}(-\omega^2 L^2+l^2)+ L& -2i\tilde Ll\cr
2i\tilde Ll & {\tilde L^2\over  L}(-\omega^2
L^2+l^2)+\cr}\pmatrix{\tilde\delta v_1\cr \tilde\delta v_2\cr}=0}
which gives the frequencies
\eqn\eeightq{\omega^2_\pm={1\over  L^2}(l\pm { L\over \tilde
L})^2={1\over  L^2}(l\pm 1)^2}

\newsec{Comments on the excitation spectrum}

In this section we discuss some  aspects of the vibration modes. We
do not discuss however the case $m=3,
n=3$ since as mentioned above there is a richer set of issues in that
case and we will discuss those details
elsewhere.

\subsec{Qualititive comments on the frequency spectrum}

At first glance one might think that the larger the radius $r_0$ of
the brane, the lower would be the frequency of its normal modes of
vibration. But we have seen that these frequencies (measured in the
coordinate $\tau$) are in fact independent of $r_0$. The reason for
such behavior can be traced to the following. Consider first the case
of branes in $AdS$.  If the graviton were pointlike it would be placed
at $r=0$, where $|g^{\tau\tau}|=1$, but because of its size the surface
of the brane is near $r=r_0$ where $|g^{\tau\tau}|\approx (1+{r^2\over
\tilde L^2})^{-1}$. Thus if we look at large $r_0$ we will have
$|g^{\tau\tau}|\sim r_0^{-2}$, which is the same as the behavor
$g^{\alpha\alpha}\sim r_0^{-2}$. Thus the frequencies, which see the
ratio of tension to density of any extended object, become independent
of $r_0$ in this limit. When we take into account the motion in the
$\phi$ direction as well in the complete analysis, then this rough
statement in fact becomes exactly true. The induced metric for the
equilibrium configuration also includes a contribution from the
angular velocity which in fact precisely cancels the $1$ and leaves us
with a result which is simply $g^{\tau\tau} =- {{\tilde L}^2 \over
r_0^2}$. Since the contribution from the angular momenta of the
fluctuations to the action
also scale as ${1\over r_0^2}$, the curvature scale of
the background determines all the frequencies of the giant
graviton. For branes in $S^n$ there is a similar effect. Now the
contribution of the angular velocity to the induced $g_{\tau\tau}$ is
$( 1 - {q_0^2 \over L^2})$ and this combines with the usual
contribution from the target space metric to lead to a final
$g^{\tau\tau} = -{L^2 \over q_0^2}$.
In fact it may be easily verified that
by a suitable rescaling of the fluctuation fields all the $r_0$ or
$q_0$ dependences can be completely scaled out of the small fluctuations
action.

\subsec{Modes arising from shift of BPS configuration}

We do not find any unstable modes in the system at this quadratic
order in the analysis, as all the $\omega^2$ are real and nonnegative.
In the formulae for $\omega^2$ the $Q_l$ are nonnegative
numbers. Further, $Q_l=0$ for $l=0$ (which is the mode constant over
the $\sigma$ coordinates), and $Q_l>0$ for $l>0$.

We can see in the spectrum the consequence of the fact that we have
several parameters that can be varied in the equilibrim configuration.
Let us examine such modes in turn.

(a) \quad From \esix\ and \esixq\ we
see that $\omega^2=0$ is one of the solutions when $Q_l$=0,  in the
$\delta r, \delta \phi$ system and in the $\delta q, \delta \phi$
system. This zero mode corresponds to the fact that the radius $r_0$
or $q_0$ of the equilibrium configuration can be taken to have any
chosen value allowed by the geometry. Different values of $r_0$ and
$q_0$ have different energies, but the same value of the action,
viz. zero.

(b)\quad Consider branes expanding in
the AdS.  We have taken the brane to move along the $\phi$ direction
at $\theta=0$ on the sphere $S^n$. But we could let the brane rotate
along some other great circle on $S^n$. We can achieve this by a
rotation
\eqn\ytwo{z'_1 =z_1 \cos\alpha - y_1\sin\alpha,
~~~~y'_1=z_1\sin\alpha + y_1\cos\alpha}
with $\alpha$ a constant.
For small $\alpha$ we find that starting with the configuration with
$z_1=\cos\theta\cos\phi=\cos\phi=\cos(\omega_0\tau)$  (see eq. \six )
and $y_1=0$ we
get a
configuration with
\eqn\ythree{z'_1\approx z_1, ~~~~y'_1\approx \alpha
\cos(\omega_0\tau)=\alpha \cos({1\over L}\tau)  }
This pertubation has  $l=0$ (since the deformation is
independent of the $\sigma_i$) and agrees with the frequency
$\omega^2=1/ L^2$ obtained from \efour\ for the $y_i$ vibrations with $l=0$.

(c) \quad By a similar analysis, if we study the branes on the sphere
and look at $v_i$ perturbations with
$l=0$ we find the frequency $\omega^2=1/\tilde L^2$, in agreement
with \efourq .  We interpret these
modes as `boosts' in the AdS space, and the frequency corresponds to
the natural period in time $t$ of this
space.

Thus we have found all the expected families of solutions: the
different values of $r_0$ or $q_0$ (corresponding to different $P_\phi$),
the different orientations of angular momenta and the possible motions
in the AdS space of the center of mass of the brane.

(Note that we have not discussed here the details of the case  $m=3, 
n=3$, where there are
additional zero modes seen at $l=\pm1$ in \eeight\ and \eeightq .)

\subsec{Excitation spectrum in a Hamiltonian analysis}

Now let us ask the question:  What would be the excitation energies
of the brane that would result from a
quantization of the small vibrations around the configuration with
$E=|P_\phi| = J$?  When there are are continuous
family of equilibrium solutions as in this case, one should do the
following. Fix the values of conserved
quantities, and for any choice of these quantities write down the
classical Hamiltonian.  If fixing the the
conserved quantities gives a unique lowest energy state, then the
Hamiltonian for small perturbations will
be a quadratic form in the coordinates and momenta describing the
perturbation. Then one does the usual
diagonalisation of quadratic forms and extracts the classical
frequencies of oscillation $\omega$. If we then
consider the quantum problem, the energy of an excited configuration will be
\eqn\uone{E\approx E_0 + \hbar\omega = J+\hbar\omega}
The value of $E_0=J$ will be itself quantised too, since the angular
momentum operator in quantum
mechanics has a discrete spectrum.

Our analysis has been Lagrangian rather than Hamiltonian, and we did
not fix the values of conserved
quantities; we found {\it all} solutions near the equilibrium
configuration. Of course if there is an oscillation
with frequency $\omega$ found from the Hamiltonian with fixed values
of the conserved quantities, then
this oscillation and its frequency will be found among our Lagrangian
solutions.  But some of the solutions
found in the Lagrangian method will not appear in the Hamiltonian
analysis, since they will not hold fixed
the values of the conserved quantities. After we locate these latter
modes and remove them from the
spectrum, we will be left with the modes that will correspond to the
excitation levels \uone .

    We will now see that the modes to be removed are precisely those
that we looked at in the last subsection.
The conserved quantities to be held fixed are all the components of
the angular momentum $J_{ab}$, and the
conserved quantities corresponding to
generators of the isometries of $AdS_m$ (which we loosely call
"momentum").
The mode of type (a) in the above
subsection leads to a change in the
magnitude of
$J$ (without changing its direction). The mode in (b) gives a nonzero
value of the angular momentum in the
$z_1-y_1$ plane, (while the original angular momentum was in the
$z_1-z_2$ plane); thus it also does not
hold the angular momentum fixed. Similarily the mode of type (c)
leads to a change of `momentum' in the
AdS space.  Thus these modes will not appear in the analysis of the
Hamiltonian with fixed values of
conserved quantities.

All the above modes had $l=0$. Note that if $l\ne 0$, then there will
be no change in the conserved quantites
when we excite the mode. This is because these conserved quantities
appear as an integral over a
$\tau=constant$ hypersurface of the worldvolume,  with the integrand
being for example the angular
momentum density. If $l\ne 0$ then we get no contribution to the
conserved quantity at linear order in the
perturbation.\foot{We may still get a contribution of order
$\epsilon^2$ to  the value of a conserved
quantity like
$J$  (under the perturbation of order $\epsilon$), but we can undo
this change by an order $\epsilon^2$
change in the equilibrium value of
$r_0$, and so can regard the perturbation as one that gives no change
in the conserved quantity.}

Thus the only modes that we could lose when working with fixed values
of the conserved quantities are
modes with $l=0$. Apart from the modes (a)-(c) of the above
subsection, there is only one such mode with
$l=0$ for the brane expanding in AdS space and one for the brane
expanding on the sphere.  Consider for
concreteness the case of the brane expanding in AdS space. From
\esix\ we see that this mode has frequency
\eqn\utwo{\omega=\pm {m-3\over \tilde L}}
But a short calculation reveals that under this mode the shift of the
value of $P_\phi=J$ is in fact zero. In this mode
the value of $r$ changes with time, but the value of $\dot\phi$
changes as well, so that the net change in
angular momentum ends up being zero. To verify this we first find the
eigenvector for this eigenvalue from
\efive ; we choose $\omega=(m-3)/\tilde L$ from \utwo\ for
concreteness. Then we get
\eqn\uthree{{\tilde \delta r\over \tilde\delta\phi}=i{L\over \tilde
L} {r_0^2+\tilde L^2\over r_0} }

But we  have for the angular momentum density on the brane
\eqn\ufour{p_\phi=T_{m-2}{L^2
r^{m-2}\dot\phi\over
\sqrt{1+{r^2\over \tilde  L^2}-L^2\dot\phi^2}}}
where we have dropped terms like $\dot r^2$
and $\dot y^2$ from the denominator that  will
vanish to linear order around the equilibrium
configuration.

Using \uthree\ we find that under the perturbation,
\eqn\ufive{\delta p_\phi={\p p_\phi\over \p r}\delta r + {\p
p_\phi\over \p \dot\phi}\delta \dot\phi =0}
Thus the angular momentum does not change, under this perturbation,
and the perturbation will survive
in the Hamiltonian analysis that gives the vibration spectrum. Due to
the symmetry between the
cases of branes on the AdS space and branes on the sphere, we expect
that a similar conclusionwill
hold for the
$l=0$ mode for branes expanding on the sphere; this mode has
$ \omega = \pm {n-3 \over L}$
from \esixq .

Thus we conclude that the vibration spectrum at fixed values of
angular momenta and AdS momenta is
given by all the modes found in the Lagrangian analysis with the
exception of those discussed in (a)-(c) in
the previous subsection.

\newsec{Discussion}

We have not found any families of BPS solutions besides the known
ones, so the ansatz using spherical branes for BPS configurations
appears to be an adequate one. The excitaion modes all have real
positive $\omega^2$, so we have not found any instabilties, for any
size of the brane. Thus this analysis does not shed light on the issue
of whether BPS branes can exist in the AdS space with arbitrarily
large angular momentum. Note however that when branes expand in the
AdS space then we use very little knowledge of the sphere in the
spacetime - we just use the motion around the equator of the
sphere. Thus if the compactification had a torus in the internal space
(as in the case $AdS_3\times S^3\times T^4$) then we could let the
brane move along a circle on the torus instead. Then the question of
whether we should have an upper bound to the brane size would be
related to whether we expect a limit on the total U(1) charge of the
graviton state.

In \dtv\ it was found that the giant graviton phenomenon occurs in
spacetimes other than $AdS \times S$. In particular, consider
$p$-branes wrapping the transverse $S^{p+2}$ of the near horizon
geometry of $D(6-p)$ branes (both extremal and near-extremal) in
string theory. In Poincare coordinates there is now an equilibrium
solution where the brane has a fixed size, carries angular momentum on
the $S^{p+2}$ and moves in the radial direction transverse to the
$D(6-p)$ brane. The energy of this system as a function of the radial
and angular momenta is exactly the same as that of a graviton moving
in this geometry. Note that generically these states are not BPS,
though they are the lowest energy states for a given angular momentum.
It would be interesting to examine the issue of small fluctuations
around these configurations.

Let us comment on a significant implication of our result. The
gravitons with high angular momentum correspond to chiral operators
with high scaling dimensions in the dual CFT: $\Delta\sim E$, where
$E$ is the energy (and angular momentum) of the graviton (measured in
units of the curvature length scale). The fact that the excitation
spectrum of this graviton has spacings of order the curvature scale
means that, if the giant graviton picture is right, these chiral
operators have a set of associated nonchiral operators with dimensions
increasing in steps of order unity. It would be interesting to look
for such a spectrum explicitly in a strongly coupled CFT, for example
in the $d=4$ super-Yang-Mills theory.

In \luninmathur\ it was found that in the orbifold CFT corresponding
to a D1-D5 system, the correlation function of three chiral primaries
drops significantly below the naive supergravity expectation when the
spin of the chiral primaries becomes comparable to the square root of
$N=Q1Q5$.  This phenomenon is a manifestation of the stringy exclusion
principle, which truncates the spectrum of chiral primaries at spin
equal to $N$. Let us assume for the moment that in the supergravity
limit of the theory, this decrease in coupling continues to be
valid. One possible explanation for the drop in coupling could be that
when the energy of two interacting supergravity quanta becomes very
high, we produce particles other than supergravity modes, and thus the
amplitude to produce just a third BPS graviton by colliding two
gravitons becomes very small.\foot{We thank E. Martinec for a
discussion on this point.}

If the giant graviton picture describes supergravity quanta at high
energy, then we see that some of the energy of interaction may in fact
go into exciting the vibration modes of the gravitons. Since the
spacing between excitations of the graviton excitations is
comparatively small, a large number of modes are available to be
excited, and the effect of exciting these modes can be quite
significant. (A typical graviton mode in AdS space has a frequency
that is of the order of the AdS
scale or higher, so interactions will generally be able to excite
these vibration modes.)

\bigskip
\bigskip

{\bf Acknowledgements}
\bigskip

We would like to thank S. Trivedi for dicsussions.  A. Jevicki was
partially supported by Department of Energy under grant
DE-FG02-91ER4068-Task A.  S.D. Mathur is supported in part by DOE
grant no. DE-FG02-91ER40690.

\listrefs
\end